\documentclass[10pt,letterpaper]{article}
\usepackage{opex3}

\begin{document}


\title{Power Scaling in High Speed Analog-to-Digital Converters using Photonic Time Stretch Technique}

\author{Shalabh~Gupta$^{1,*}$, George~C.~Valley$^2$, Robert~H.~Walden$^2$, and~Bahram~Jalali$^1$}

\address{$^1$Department of Electrical Engineering, University of California, Los
Angeles, CA 90095}
\address{$^2$The Aerospace Corporation, El Segundo, CA 90245}

\email{$^*$shalabh@ee.ucla.edu} 



\begin{abstract}
Factors that contribute to the rapid increase in power dissipation as a function of input bandwidth in high speed electronic Analog-to-Digital Converters (ADCs) are discussed. We find that the figure of merit (FOM), defined as the energy required per conversion step, increases linearly with bandwidth for high-speed ADCs with moderate to high resolution, or equivalently, the power dissipation increases quadratically.  It is shown that by use of photonic time-stretch technique, it is possible to have ADCs in which this FOM remains constant for up to 10 GHz input RF frequency. Using this technique, it is also possible to overcome the barrier to achieving high resolution caused by clock jitter and speed limitations of electronics in such ADCs.  

Use of optics is actively being pursued for reducing power dissipation and achieving higher data-rates for board-level and chip-level serial communication links. In the same manner, we expect that optics will also help in reducing power dissipation in high-speed ADCs in addition to providing broader bandwidths. 
\end{abstract}



\section{Introduction}
Increased bandwidth demands from internet backbones have led researchers to target 100 Gbit/s
and higher data rates per wavelength division multiplexing (WDM) channel using spectrally efficient,
multilevel modulation formats \cite{WinzerLeos2007}. Demodulation of such signals requires analog-to-digital converters (ADCs) with very high performance and resolution. Such ADCs are also crucial for defense applications such as radars and for wide-bandwidth laboratory instruments such as oscilloscopes and vector spectrum analyzers. 

While continued scaling of CMOS technology \cite{Muller2005} has improved digital circuits tremendously in terms of performance, power efficiency and cost, analog circuits (and ADCs) have not really kept pace. Even though the bandwidth of an analog circuit improves with technology scaling, since smaller devices run faster, thanks to reduced capacitances, power dissipation for the same functionality does not always scale because of lower intrinsic gains in shorter channel CMOS transistors \cite{Annema1999}. In fact, most improvements in power efficiency of analog circuits can be attributed to architectural improvements and scaled voltages, rather than reduced capacitances in CMOS devices. 

\begin{figure}[!t]
\centering
\includegraphics[width=3.4in]{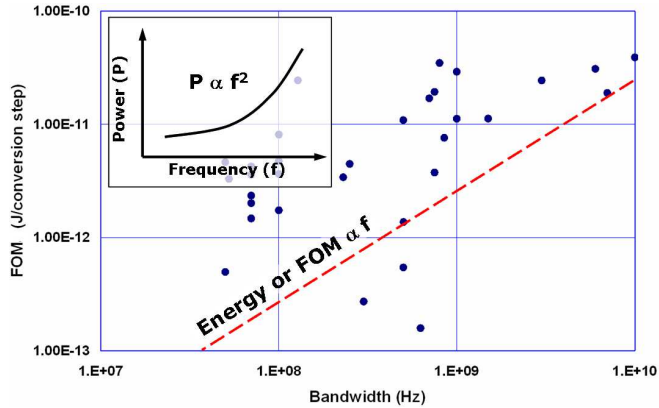}
\caption{Energy per conversion step versus performance for  state-of-the-art high-speed analog-to-digital converters with moderate to high resolution (from 2002 to present). FOM roughly increases linearly, i.e., power dissipation increases quadratically with frequency (i.e. $P \propto f^2$, as shown in the inset).}
\label{trends} 
\end{figure}

In ADCs, analog full-scale voltage cannot be reduced arbitrarily because of the thermal ($kT/C$) noise limitations. As a result, while technology scaling has not directly resulted in decreased analog power dissipation, power reduction and increased speed of digital circuits has allowed extensive use of digital correction and calibration techniques, which have led to improvement in performance. The most commonly used figure of merit (FOM) for ADC efficiency is the energy required per conversion step, 

\begin{equation}
FOM = \frac{P_{diss}}{2^{ENOB}\times2\times ERBW},
\label{Eq1}
\end{equation}               
  
where, $P_{diss}$ is the power dissipation of the ADC, ENOB is the effective number of bits and ERBW is the effective resolution bandwidth of the ADC \cite{Walden1999}.  ERBW is defined as the frequency at which the signal-to-noise-and-distortion ratio (SNDR) of the ADC degrades by 3-dB as compared to its low frequency value. Fig. \ref{trends} shows the energy per conversion step as a function of ERBW for ADCs from 2002 to the present.  While Fig. \ref{trends} lumps together ADCs for 5 or 6 different electronic technologies, it is evident from the plot that above 100-MHz, energy per conversion step for the best ADCs increases rapidly with bandwidth. Similar trend is found from the data obtained in \cite{MurmannOnline} for CMOS technologies for high speed ($>$100MHz ERBW) and moderate to high ($>$6 ENOB) resolution ADCs. In reference \cite{MurmannCICC2008}, Murmann finds that power the efficiency of the ADCs has improved by a factor of 2 every two years, thanks to technology and power supply scaling. However, the data in \cite{MurmannOnline} confirms that this trend is not observed for high speed ADCs with moderate to high resolution. 

\begin{figure}[!t]
\centering
\includegraphics[width=4.3in]{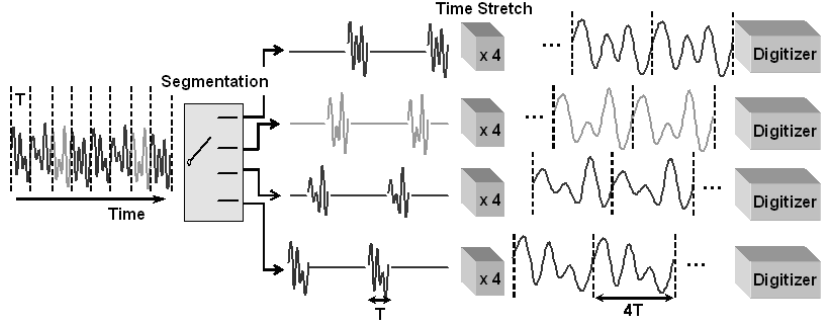}
\caption{Conceptual diagram of a photonic Time-Stretch Analog-to-Digital Converter with a stretch factor of 4. The high-speed RF signal to be digitized is segmented into multiple wavelength channels and time. All segments are then time stretched and digitized by much slower backend digitizers. Digitized signals are rearranged and combined digitally to obtain the digital representation of the original RF signal.}
\label{cont_tsadc} 
\end{figure}

In this paper we estimate that the power dissipation in such high speed ADCs increases approximately as $f_s^2$ for a constant resolution, where $f_s$ is the sampling frequency with Nyquist rate sampling. On the other hand, if the photonic time-stretch technology is used \cite{Bhushan1998, Jalali2001, HanJLT2003}, we show that one can obtain linear power scaling of future ADCs to much higher frequencies. The conceptual diagram of a Time-Stretch Analog-to-Digital Converter (TS-ADC) is shown in Fig. \ref{cont_tsadc}. The time-stretch technique also allows breaking through the so-called walls in the ENOB-bandwidth plane caused by comparator ambiguity and timing jitter \cite{Walden1999, Walden2006}.    

In this paper, first we discuss the fundamental limits on noise and power dissipation in ADCs. Second, we discuss frequency scaling in digital circuits and show why similar trends are observed in ADCs. Third, we show that the photonic time-stretch technique can push the linear scaling of power dissipation versus sampling frequency in the ADCs well into the GHz band. Finally, we compare the power dissipation in high speed ADCs with and without the use of the photonic time-stretch technique.

\subsection{Fundamental limits to power dissipation in ADCs}
An analog-to-digital converter consists of two main operational stages.  First, the sample and hold stage (S\&H) samples the analog signal on a capacitor through a switch at periodic intervals of time. Second, the quantization stage which converts the sampled analog signal using comparators to digital (binary) outputs.  Depending on the architecture, there can be a series of these two stages in a specific ADC design.  In addition, clock generation circuitry is required to provide clocks with different phases and duty cycles to different sets of switches so that all operations in the ADC are synchronized. Finally, there are reference buffers that act as accurate voltage sources with very low output impedances. Additionally, modern ADCs use digital circuitry extensively for correction and calibration in post-processing, which can also consume a significant fraction of the total power. 

In the sample and hold stage, the analog signal and the thermal noise generated in the switch are sampled onto a capacitor with capacitance $C$. The sampled noise voltage has a mean squared value of $kT/C$, where, $k$ is the Boltzmann constant and $T$ is the ambient temperature.  Note that the total magnitude of thermal noise is sampling frequency independent.  This is because the thermal noise at all frequencies is folded back into the Nyquist bandwidth in a sampled system. If the thermal noise is the dominant noise source and $V_{FS}$ is the full scale voltage, we obtain the signal-to-noise ratio and dissipated power as:

\begin{eqnarray}
SNR \propto \frac{V_{FS}^2}{kT/C}, ~~~~~P_{diss} \propto CV_{FS}^2f_s \\
\Rightarrow ~~~P_{diss} \propto kT.f_s.SNR~~~~~~~~~~~~~~~
\label{Eq2}
\end{eqnarray}

Therefore, for a given signal power (or $V_{FS}$), thermal ($kT/C$) noise places a lower limit on the capacitance that can be used in the sample and hold stage. On an average, bias currents required in the buffer amplifiers to charge the sampling capacitors with the signal or the reference voltage are directly proportional to the value of $C$ and the charging time, fundamentally limiting the power dissipation of an ADC, as observed in (\ref{Eq2}). This limitation can be written in terms of the effective number of bits of the ADC by substituting the standard relation for ENOB as a function of SNR \cite{Walden1999,Kenington2000}: 

\begin{equation}
P_{diss} \propto kT.f_s.10^{(6.02\times ENOB+1.76)/10}.
\label{Eq3}
\end{equation}

In practical circuits, the power dissipation is at least 3 to 4 orders of magnitudes higher than this value \cite{Cho1994} since the individual components such as voltage buffers, opamps, comparators, and clock sources have to satisfy requirements  of low noise, high linearity, high speed and high precision settling. Currents in all these circuits scale proportionally with the capacitance $C$ for a fixed sampling frequency. 

Equation (\ref{Eq3}) suggests that ADC power dissipation should scale linearly with $f_s$. However, from the trend seen in Fig. \ref{trends} and in \cite{Poulton2006}, and from the discussion in the following subsection, it becomes clear that in reality, the energy per conversion step scales roughly as $f_s$, i.e., power dissipation is proportional to $f_s^2$ in high-speed ADCs.  

\subsection{Power Scaling in Digital Circuits}
 Power dissipation in digital circuits, which until recently has been dominated by the dynamic power required for switching transistors, is proportional to $CV^2f$, where $C$ is the average node capacitance, $V$ is the supply voltage and $f$ is the clock frequency \cite{Horowitz1994}. To run the circuits at fast speeds, high switching currents are required to charge or discharge the node capacitances quickly, which demands a higher supply voltage.  The minimum operating voltage $V$ at which a digital circuit can operate correctly is roughly proportional to $\sqrt{f}$ for a wide range of frequencies or voltages \cite{Horowitz1994}. This implies that if the frequency of operation is reduced by a factor $\alpha$, the required power decreases by factor $\alpha^2$. As a result, the energy-delay product for performing an operation is roughly constant over a wide range of operating frequencies in digital circuits \cite{Horowitz1994}. This simple observation implies that when more delay is allowed for a set of operations, less energy is required to perform them. This implication is the reason that the digital world is moving towards architectures exploiting parallelism \cite{ITRS2005},  and the same trend is found in high sample rate ADCs and real-time digital oscilloscopes \cite{Dyer1998, Poulton2003, GuptaISSCC2006, TEKOSC, LECROYOSC}. In this paper, we find that the time-stretch technique, which uses the same approach of parallelism to digitize very high bandwidth signals, can also help in reducing power dissipation in high speed ADCs. 
  
\subsection{Power Scaling in Analog-to-Digital Converters}

\emph{Speed considerations:} In deriving the expression $P_{diss} \propto kT.f_s.SNR$, it was assumed that to increase the sampling frequency, the bias currents need to be increased linearly to charge up the capacitors fast with no limitation being posed by the transistor response time. In reality, the unity gain frequency $f_T$ of the transistors should also be increased linearly with $f_s$ because the operational amplifier (opamp) outputs driving the capacitors have shorter time to stabilize before quantization begins.

Increasing the bias current in a CMOS transistor can be done in two ways. In the first method, the device size is kept constant and the bias current ($I_D$) is increased by increasing overdrive voltage ($V_{OD}$). This increases the unity gain frequency $f_T \propto \sqrt{I_D}$, as for a device with a fixed size, transconductance $g_m \propto \sqrt{I_D}$ and $f_T \propto g_m$. However it also results in a lower output resistance $r_o$ which is proportional to $1/I_D$ and hence a lower intrinsic gain of $g_mr_o \propto \sqrt{1/I_D}$. Lower gain results in higher gain errors in the opamps reducing accuracy of the ADC. 

In the second method, overdrive voltages are kept constant and only the transistor widths are increased linearly for a proportional increase in current. In this case, the intrinsic gain is maintained, but $f_T$ does not increase, resulting in an incomplete settling of the opamp outputs for higher sampling frequencies. 

In both cases, we find that just by scaling currents proportionally, one cannot fulfill the requirements of faster response time while maintaining the same linearity (and resolution). For quantization process, the voltage comparators also need to switch faster to avoid comparator ambiguity \cite{Walden1999}. The same arguments, as discussed above, indicate that increasing drive currents linearly with frequency in comparators is again not a solution for achieving the required comparator speeds. These facts suggest that power dissipation in ADCs should increase more rapidly with frequency, following a somewhat similar trend as in digital circuits. This frequency scaling trend, as shown in Fig. \ref{trends}, is found not only in CMOS technologies, but is also central to other technologies like SiGe, GaAs and InP which have traditionally been used for very high speed ADCs.

\begin{figure}[!t]
\centering
\includegraphics[width=3.4in]{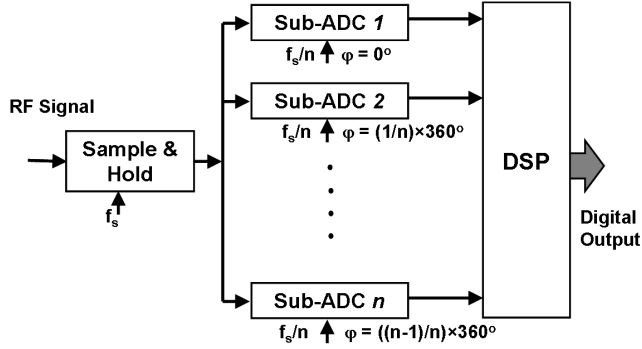}
\caption{Schematic of a time interleaved ADC.}
\label{tint_adc} 
\end{figure}

To overcome these issues in frequency scaling, new ADC architectures employing parallelism must be used as in case of digital circuits \cite{Horowitz1994}. In the time-interleaved architecture, which is a parallel ADC architecture such as the one shown in Fig. \ref{tint_adc}, multiple sub-sampling ADCs are used in parallel, to sample the signal at different instants of time within a full sampling clock cycle \cite{Dyer1998,Poulton2003,GuptaISSCC2006}. The outputs of these ``sub-ADCs" are combined in the digital domain and post processing is performed to suppress distortions caused by timing errors, gain mismatches and DC offsets. The front-end of the time interleaved architecture can have a single S\&H block \cite{Dyer1998} feeding all sub-ADCs, or a separate S\&H block corresponding to each sub-ADC \cite{Poulton2003}, or a combination of the these two approaches \cite{GuptaISSCC2006}. In the first and the third case, scalability to high sampling frequencies and to large sub-ADC numbers is still a challenge, and same power considerations, as discussed above for the ADCs, apply to the front-end circuitry. The second architecture (discussed in \cite{Poulton2003}) can potentially be scaled to have higher sampling rates, but timing jitter and residual timing offsets, which are discussed in the next section, limit the ADC resolution.  Also, this architecture requires a predriver to drive a large capacitive load of S\&H blocks which can limit the bandwidth and add significant power dissipation.

\subsection{Noise due to Aperture Jitter}

\begin{figure}[!t]
\centering
\includegraphics[width=4.3in]{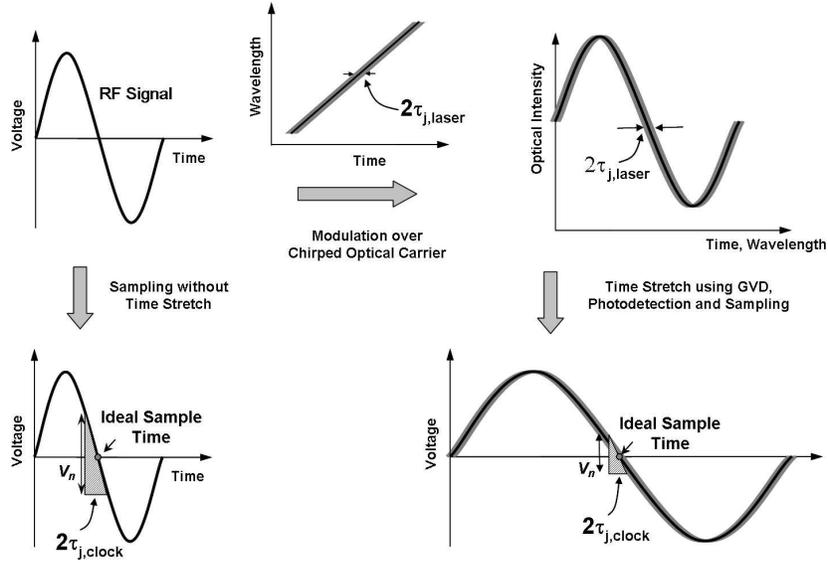}
\caption{Sampling a signal with and without time-stretch. When time-stretch technique is used, the noise due to clock jitter becomes insignificant, and noise added by the laser jitter dominates, which is typically much less than the electronic clock jitter ($V_n$ represents the noise voltage).}
\label{jitters} 
\end{figure}

In high speed ADCs, aperture jitter (or uncertainty) is a significant source of noise \cite{Walden1999} (as shown in Fig. \ref{jitters}), which is caused by jitter in the sampling clock. Aperture jitter noise is signal frequency dependent which severely degrades the SNR of moderate to high frequency signals. The jitter limited SNR for an rms timing jitter $\tau_j$ is given by \cite{Walden1999, BartolomeTI2005}:

\begin{equation}
SNR_{jitter} = -20log(2\pi\tau_jf_{signal}).
\label{Eq4}
\end{equation}

Most of the aperture jitter noise is added by the jitter in the sampling clocks generated by the clock sources. In the time interleaved architecture used in \cite{Poulton2003}, timing errors in clocking different sub-ADCs also add the same effect as jitter and limit the achievable SNR. For example, the 20 GS/s ADC reported in \cite{Poulton2003} shows a resolution of 6.5 effective bits at low frequencies, but the resolution drops to 4.6 effective bits for 6-GHz RF signal because of an effective rms sampling jitter of 0.7-ps. In another example \cite{BartolomeTI2005}, an rms jitter of 250-fs is shown to reduce the effective resolution of a 14-bit ADC to about 11.2 effective bits for a 230-MHz RF signal. As discussed in the next section, the time-stretch ADC technique uses optical processing to overcome these limitations, in addition to achieving significant power savings.

\section{Time Stretch Analog-to-Digital Converter}

\begin{figure}[!t]
\centering
\includegraphics[width=3.4in]{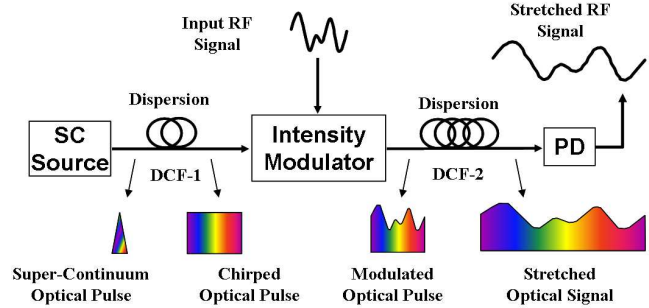}
\caption{Schematic diagram showing physics of photonic time-stretch preprocessing technique. The RF signal is modulated over a linearly chirped optical pulse obtained by dispersing a super-continuum (SC) pulse. Signal obtained at the photo-detector output, after propagation through the second dispersion fiber, is a time-stretched replica of the original RF signal.}
\label{tsadc_fund} 
\end{figure}

In a time-stretch ADC (TS-ADC), the effective bandwidth and frequency of the RF signal to be digitized is compressed by stretching the signal in time \cite{Bhushan1998, Jalali2001, HanJLT2003}, thereby reducing the bandwidth of the backend electronic digitizer required to capture the original signal. Fig. \ref{tsadc_fund} shows the fundamental process of time-stretch. To do so, the RF signal is modulated over a long pulse of a linearly chirped optical carrier obtained from a super-continuum source (which can be a femto-second mode locked fiber laser). Propagation through a dispersive medium stretches the modulated pulse in time, resulting in a ``time-stretched" replica of the original RF signal after photodetection. The magnification or the stretch ratio $M$ is given by $(D_2/D_1 + 1)$, where $D_1$ and $D_2$ are the dispersion values of dispersion fibers DCF-1 and DCF-2, respectively. To achieve continuous operation, the optical spectrum is segmented into multiple channels using a wavelength division multiplexing (WDM) filter. Time-stretched signals from different channels are digitized by separate electronic digitizers and combined together in digital domain. Fig. \ref{cont_tsadc} illustrates one realization of the TS-ADC system that can be used to stretch the signal by up to a factor of four, and requires four channels to capture the whole signal continually.

As illustrated in Fig. \ref{jitters}, stretching the signal in time using optical preprocessing reduces the effective signal frequency seen by the S\&H block by the stretch factor $M$.  As a result, the noise added to the system due to clock jitter is scaled down by $M^2$.  For example, if a stretch factor of 10 is used in a TS-ADC, the clock jitter limited noise can be lowered by up to 20-dB compared to a conventional ADC. We note that the timing jitter of the mode-locked fiber laser, which is used for generating chirped pulses in the TS-ADC, still adds noise, but it can be reduced to very small values with careful design.  For example, a laser with 18-fs rms jitter has been reported in \cite{Chen2007}. On the other hand, the best jitter performance achieved  by clocks in electronic digitizers is of the order of 200-fs \cite{BartolomeTI2005}. In reference \cite{Zanchi2005}, the best clock jitter of 180-fs is observed for clocks with very high voltage swings. However, such voltage swings at high speeds add very substantially to power dissipation, and make clock distribution almost impossible in time-interleaved ADCs.

In time-interleaved ADCs, the clock jitter is generally much higher as extensive clock generation circuitry is required to generate multiple clocks with very precise phase delays. Additionally, even after adaptive alignment and calibration, there is a residual timing misalignment between clocks going to different sub-ADCs.  Stringent requirements on clock accuracy can thus result in  a significant power penalty in the time interleaved ADCs. 

As evident from Fig. \ref{jitters}, when the time-stretch technique is used, the effective sampling jitter in the system can be reduced, and can be written as 

\begin{equation}
\tau_{j, effective} = \sqrt{\tau_{j, laser}^2 + (\tau_{j, clock}^2/M)^2}.
\label{Eq5}
\end{equation}

This makes the time stretch architecture very well suited for high signal frequency applications where aperture jitter is the dominant source of noise. Another key advantage of time stretching is that none of the electronic digitizers see the original, very high frequency signal since the signal frequency scales down upon stretching. As a result, the power scales linearly with  sampling frequency for the ADC. In the next sub-section, we consider an example to show how time stretching can be very useful in the context of power savings. 

\subsection{Power calculations for a Time-Stretch ADC}

\begin{figure}[!t]
\centering
\includegraphics[width=3.4in]{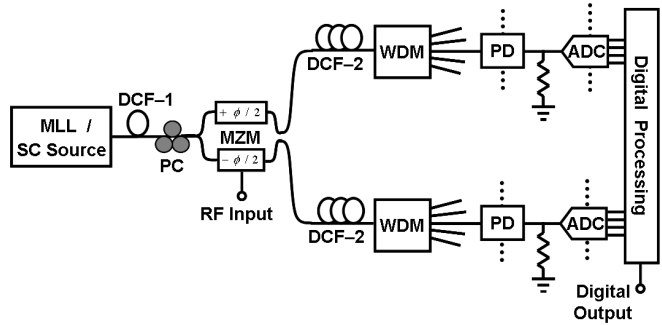}
\caption{Schematic diagram of the Time-Stretch ADC for continuous operation (MLL: Mode Locked Laser; SC: Super-Continuum; PC: Polarization Controller; DCF: Dispersion Compensating Fiber; MZM: Mach-Zehnder Modulator; PD: Photodetector).}
\label{tsadc_blkdiag} 
\end{figure}

The block diagram of a TS-ADC system for continuous operation is shown in Fig. \ref{tsadc_blkdiag}. We assume that the repetition rate of optical pulses from the mode-locked laser (MLL) is 100-MHz. Therefore, the time segments that need to be captured by electronic digitizers are 10-ns long. Usable optical bandwidth of 40-nm (i.e. 5-THz bandwidth in frequency at 1550-nm center wavelength) can easily be obtained from a femto-second MLL, for example, the FFL1560-MP laser from Precision Photonics, followed by a highly non-linear fiber \cite{GuptaOL2008}. For continuous modulation of RF signal, these 40-nm pulses have to be stretched to 10-ns before they are modulated using Mach-Zehnder modulator (MZM), which requires -250 ps/nm dispersion in DCF-1, corresponding to dispersion of about 15-km SSMF (standard single-mode fiber).  For the TS-ADC, dispersion compensation fibers (DCFs) are used because they have higher dispersion-to-loss ratios compared to SSMF. Using a DCF, dispersion value of -250ps/nm is achieved with a distributed loss of about 0.65-dB (as found in \cite{OFSDCF} and from measurements in our lab), to which connector losses are added separately.  If the stretch factor is $M$, the dispersion required in DCF-2 becomes $(M-1)\times(-250ps/nm)$, since $M = (D_2/D_1 + 1)$, where $D_1$ and $D_2$ are the dispersion values of DCF-1 and DCF-2, respectively. Also, we estimate the losses in the Mach-Zehnder modulator to be 4-dB and losses in WDM filter, polarization controller and connectors to be an additional 3-dB. Therefore, total loss in the optical link is about $M\times0.65 + 4 + 3 = (M\times0.65 + 7) dB$.  Also we estimate that the power at the input of each photodetector as 1 mW -- which gives 58-dB shot noise limited SNR for 500-MHz RF bandwidth, 0.5 modulation depth and 0.8 A/W photodetector responsivity. The noise and SNR calculations for an optical system is shown in Appendix A. Same or better thermal noise from electronics and photodetector can easily be achieved, resulting in better than 58-dB SNR with differential operation \cite{GuptaJLT2007}. Backend ADCs with 8-ENOB (i.e. 50-dB SNDR) can now be used to obtain same 8-bit resolution, as the additional noise is compensated by differential operation \cite{GuptaJLT2007}. Noise contribution due to laser RIN (relative intensity noise) is unimportant as a modest RIN of -150-dB yields an SNR of 63-dB in such conditions.

\begin{figure}[!t]
\centering
\includegraphics[width=3.4in]{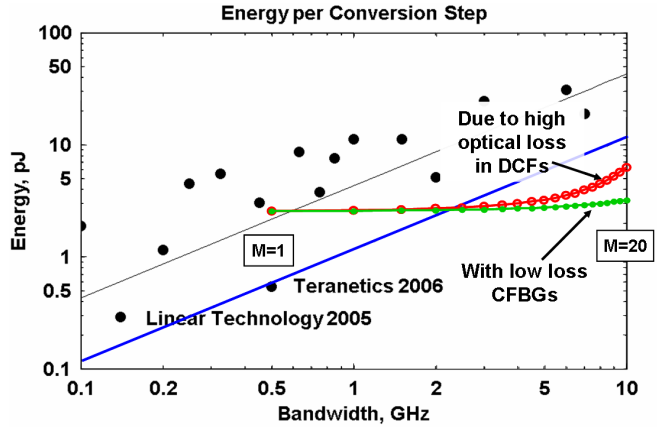}
\caption{Projected FOM for a Time-Stretch ADC (red trace) and a time-interleaved ADC (blue line) for different signal frequencies. Green trace shows FOM if CFBGs are used for dispersion, which also represents a trend if optical amplification were used.}
\label{tsadc_power} 
\end{figure}

The backend digitizers are assumed to capture waveforms at a sample rate of 1-GS/s, with 0.5-pJ/step FOM and 500-MHz Nyquist bandwidth with 8-ENOB -- resulting in 125-mW power dissipation. These values are projected using the blue line in Fig. \ref{tsadc_power} obtained from observing the linear dependence of ADC FOM on $f_s$ for published ADCs, and using FOM of the best reported GS/s ADC \cite{GuptaISSCC2006}. For each optical WDM channel, \emph{differential} and \emph{arcsine} operations are performed \cite{GuptaJLT2007, JuodawlkisMTT2001}, which not only improve the SNR by 3-dB but also suppress non-linear distortion due to electro-optic modulation and chromatic dispersion. However, this requires that the number of backend digitizers and photo-detectors are twice the stretch factor $M$ (or the number of optical channels). The electrical-to-optical power conversion efficiency of the laser is assumed to be 20\% and the power consumed by each photodetector is estimated to be 50-mW. This includes the power required to bias the photodetector and the power in the subsequent amplifying stage to bring the signal to full scale voltage of the electronic digitizer. As a proof of principle, a two channel 7-ENOB TS-ADC with 10-GHz RF bandwidth was recently demonstrated \cite{GuptaOL2008}, in which the resolution was primarily limited by the backend digitizer. This is, to the best knowledge of the authors, a \emph{world record} resolution achieved in digitization of 10-GHz bandwidth signals.

Finally, the combination of channel outputs in digital domain requires signal processing and memory.  Even though CMOS scaling has made digital circuits highly power efficient, large amount of digital data is generated, which requires significant power consumption in digital post-processing. Digital power is estimated to be the same as the total power consumed in backend electronic ADCs (as a similar trend is observed in \cite{Poulton2006}). Using these numbers with 1-mW input optical power at each photo-detector, the total optical and electrical powers can be calculated. Power scaling obtained in the TS-ADC is plotted as the red curve in Fig. \ref{tsadc_power}.  It is observed that the FOM roughly stays constant up to 5-GHz as power consumption of electronics dominates at lower frequencies.

For higher frequency signals, longer dispersive fibers are required to have larger stretch ratios, which add significant power penalty due to optical losses. Optical amplification using Erbium doped fiber amplifiers, or distributed Raman amplification can be used to curtail these losses and  improve the overall power efficiency significantly while maintaining high SNR. However, their discussion has been avoided here for simplicity. Furthermore, with lower loss dispersive media, such as chirped fiber Bragg gratings (CFBGs), linear power scaling trend can continue to much larger bandwidths, as shown by the green curve in Fig. \ref{tsadc_power}. In these calculations, the losses in CFBGs are assumed to be half of the DCFs, though in actual, the CFBG losses are even lower \cite{Ouellette1994}. However, the CFBGs can have a significant group delay ripple, which must be reduced or corrected in high resolution applications. Since there is practically no fundamental limit to obtaining the lower group delay ripples in CFBGs, they hold promise for future TS-ADC applications. Moreover, distortions added by CFBG ripples are static and can be calibrated out \cite{Conway2008}. 

Table \ref{table1} summarizes the breakdown of power consumption in a 10-GS/s TS-ADC using DCFs and 20-GS/s TS-ADCs using DCFs and CFBGs as dispersive media. All these power calculations are for 8-effective bits of resolution and Nyquist rate sampling. The 10-GS/s TS-ADC is projected to consume power of about 8.3 W and the 20-GS/s TS-ADC using DCFs is expected to consume about 32 W of power. In the 20-GS/s ADC, optical power requirement increases because of longer DCFs which add more losses. However, if optical amplification or CFBGs are used, the power consumption can be reduced to about 16.5 W. Compared with this, the projected power consumption of purely electronic 10-GS/s and 20-GS/s ADCs is 12.5 W and 50 W respectively. However, at present, there are no electronic ADCs with 8-effective bits of resolution, and 10 or 20-GS/s Nyquist sampling rates -- a direct consequence of the aperture jitter limitation. 

\begin{table}[!t]
\renewcommand{\arraystretch}{1.3}
\caption{Power Dissipation Breakdown in the TS-ADC with 8-ENOB Resolution [Watts].
(DCF: Dispersion Compensation Fiber; CFBG: Chirped Fiber Bragg Grating.) }
\label{table_example}
\centering
\begin{tabular}{|p{3.0cm}|*{3}{p{2.4cm}|}}
\hline
~ & \bf{10-GS/s ADC using~DCFs (M=10)} & \bf{20-GS/s ADC using~DCFs (M=20)} & \bf{20-GS/s ADC using~CFBGs (M=20)}\\
\hline
\bf{Laser} & 2.3 & 20 & 4.5 \\ \hline
\bf{Photo-detectors}	& 1 &	2	& 2 \\ \hline
\bf{Backend Electronic ADCs}	& 2.5	& 5	& 5 \\ \hline
\bf{Digital Electronics} \& Memory	& 2.5	& 5	& 5 \\ \hline
\bf{Total TS-ADC Power} & \bf{8.3} & \bf{32}	& \bf{16.5} \\ \hline
\bf{Electronic~ADC Power for same Performance} & \bf{12.5} & \bf{50} & \bf{50} \\ 
\hline
\end{tabular}
\label{table1}
\end{table}

If a backend electronic digitizer with higher bandwidth is available, the laser pulse repetition rate in the TS-ADC can be increased, resulting in a reduced time aperture (or inter-pulse period).  Reducing time aperture reduces the dispersion values required in the system, which helps in lowering optical losses and scaling TS-ADC power linearly to even higher frequency ranges. Use of wider optical bandwidths also reduces the required dispersion and curtails optical losses. Although Mach-Zehnder modulators generally add significant wavelength dependent bias offsets for wide optical bandwidths, digital processing can easily suppress the distortions added by these bias offsets in the TS-ADC \cite{GuptaJLT2007, Stigwall2007}.

\section{Conclusion}
In this paper, we showed that the photonic time-stretch technique can be used to scale electronic ADCs to higher frequencies both by reducing power dissipation and by overcoming the SNR barrier added by electronic clock jitter and the limited speed of electronics. In addition, use of optics provides several other advantages. Optics has traditionally been very useful in transmitting very wide bandwidth analog and digital signals over large distances with low losses. In particular, transmission over optical fibers is widely used for routing analog signals from antennas at remote locations to base stations for signal processing. The TS-ADC can also be very useful in this aspect, since no additional hardware is required to provide the option of remoting in communications and radar systems. In this mode of operation, the dispersive fiber that stretches the RF signal also serves as a fiber link. 

Optical subsystems have been proposed for signal transmission at board levels \cite{ChoJLT2004} and chip levels \cite{McFadden2006} to reduce power dissipation and achieve high throughput rates. In light of this ongoing opto-electronic integration, we believe that the photonic time stretch technology has also become very important, and can be integrated with CMOS technology in near future.  

\section*{Appendix A: Noise contributed by different optical frontend components}
First we define the parameters used in the equations:

\begin{tabular}{p{1.6cm} p{9.6cm}}
~~~~~~$P_{in} =$ & Average optical power at photodetector input \\
~~~~~~$m =$ & Amplitude modulation index \\
~~~~~~$B_e =$ & Electrical bandwidth after stretching \\
~~~~~~$R =$ & Electrical impedance (50--ohm) \\
~~~~~~$\eta =$ & Photodetector responsivity \\
~~~~~~$i_n =$ & Rms noise current \\
~~~~~~$q =$ & Electron charge \\
~~~~~~$T =$ & Ambient temperature \\
~~~~~~$k =$ & Boltzmann's constant \\
~~~~~~$RIN =$ & Relative intensity noise of the laser in decibels\\
~~~~~~$P_{NEP} =$ & Photodetector noise equivalent power (typically $\sim$15-$pW/\sqrt{Hz}$).\\
\end{tabular}

~

Because of quantum nature of light, the photodetector generates shot noise current with its variance given by
\begin{equation}
i^2_{n,shot} = 2q\eta P_{in}B_e.
\label{i_shot}
\end{equation}

The noise contribution due to laser relative intensity noise (RIN) in differential operation is  
\begin{equation}
i^2_{n,RIN} = \frac{m^2}{2}(\eta P_{in})^2 10^{RIN/10}B_e.
\label{i_RIN}
\end{equation} 

Thermal noise contribution of the photodetectors is given by
\begin{equation}
i^2_{n,thermal} = 4kTB_e/R + (\eta P_{NEP})^2B_e.
\label{i_thermal}
\end{equation} 

With these three major noise contributors, and differential signaling, the total signal-to-noise (SNR) ratio in the stretched RF signal received from the time-stretch preprocessor is obtained as:
\begin{equation}
SNR_{total} = \frac{i^2_{signal}}{i^2_{n,total}} = \frac{(m^2/2)(\eta P_{in})^2}{\frac{1}{2}i^2_{n,shot}+i^2_{n,RIN}+\frac{1}{2}i^2_{n,thermal}}.
\label{snr_total}
\end{equation} 

\section*{Acknowledgment}
This work was supported by DARPA under SSC San Diego grant No. N66001-07-1-2007. The authors thank Dr. John P. Hurrell of the Aerospace Corporation for helpful comments.


\begin{thebibliography}{10}
\newcommand{\enquote}[1]{``#1''}
\expandafter\ifx\csname url\endcsname\relax
  \def\url#1{\texttt{#1}}\fi
\expandafter\ifx\csname urlprefix\endcsname\relax\def\urlprefix{URL }\fi
\providecommand{\eprint}[2][]{\url{#2}}

\bibitem{WinzerLeos2007}
P.~J. Winzer and G.~Raybon, \enquote{100G Ethernet A Review of Serial Transport
  Options,} 2007 Digest of the IEEE/LEOS Summer Topical Meetings pp. 7--8
  (2007).

\bibitem{Muller2005}
D.~A. Muller, \enquote{A sound barrier for silicon?} Nat Mater \textbf{4}(9),
  645--647 (2005). \urlprefix\url{http://dx.doi.org/10.1038/nmat1466.}

\bibitem{Annema1999}
A.-J. Annema, \enquote{Analog circuit performance and process scaling,} IEEE
  Transactions on Circuits and Systems II: Analog and Digital Signal Processing
  \textbf{46}(6), 711--725 (1999).

\bibitem{Walden1999}
R.~H. Walden, \enquote{Analog-to-digital converter survey and analysis,} IEEE
  Journal on Selected Areas in Communications \textbf{17}(4), 539--550 (1999).

\bibitem{MurmannOnline}
B.~Murmann, \enquote{ADC Performance Survey 1997-2008,}
  \urlprefix\url{http://www.stanford.edu/~murmann/adcsurvey.html.}

\bibitem{MurmannCICC2008}
B.~Murmann, \enquote{A/D Converter Trends: Power Dissipation, Scaling and
  Digitally Assisted Architectures,} in \emph{Proc. IEEE Custom Integrated
  Circuits Conference (CICC)}, pp. 105--112 (2008).

\bibitem{Bhushan1998}
A.~S. Bhushan, F.~Coppinger, and B.~Jalali, \enquote{Time-stretched
  analogue-to-digital conversion,} Electronics Letters \textbf{34}(9), 839--841
  (1998).

\bibitem{Jalali2001}
B.~Jalali and F.~Coppinger, \enquote{Data conversion using time manipulation,}
  US Patent (6288659) (2001).
  \urlprefix\url{http://www.freepatentsonline.com/6288659.html}.

\bibitem{HanJLT2003}
Y.~Han and B.~Jalali, \enquote{Photonic Time-Stretched Analog-to-Digital
  Converter: Fundamental Concepts and Practical Considerations,} J. Lightwave
  Technol. \textbf{21}(12), 3085 (2003).
  \urlprefix\url{http://jlt.osa.org/abstract.cfm?URI=JLT-21-12-3085}.

\bibitem{Walden2006}
R.~H. Walden, \enquote{Analog-to-Digital Conversion in the Early 21st Century,}
  in \emph{IEEE MTT Workshop (WMK) on Ultrafast Analog-to-Digital (A/D)
  Conversion Techniques and its Applications}, 5 (2007).

\bibitem{Kenington2000}
P.~Kenington and L.~Astier, \enquote{Power consumption of A/D converters for
  software radio applications,} IEEE Transactions on Vehicular Technology
  \textbf{49}(2), 643--650 (2000).

\bibitem{Cho1994}
T.~Cho, D.~Cline, C.~Conroy, and P.~Gray, \enquote{Design considerations for
  low-power, high-speed CMOS analog/digital converters,} in \emph{IEEE
  Symposium on Low Power Electronics Digest}, pp. 70--73 (1994).

\bibitem{Poulton2006}
K.~Poulton, R.~Neff, B.~Setterberg, B.~Wuppermann, and T.~Kopley,
  \enquote{Architectures and Issues for Gigasample/second ADCs,} Springer
  Netherlands pp. 17--32 (2006).

\bibitem{Horowitz1994}
M.~Horowitz, T.~Indermaur, and R.~Gonzalez, \enquote{Low-power digital design,}
  in \emph{IEEE Symposium on Low Power Electronics Digest}, pp. 8--11 (1994).

\bibitem{ITRS2005}
\enquote{International Technology Roadmap for Semiconductors,}  (2005).
  \urlprefix\url{http://www.itrs.net.}

\bibitem{Dyer1998}
K.~C. Dyer, D.~Fu, S.~H. Lewis, and P.~J. Hurst, \enquote{An analog background
  calibration technique for time-interleaved analog-to-digital converters,}
  IEEE Journal of Solid-State Circuits \textbf{33}(12), 1912--1919 (1998).

\bibitem{Poulton2003}
K.~Poulton, R.~Neff, B.~Setterberg, B.~Wuppermann, T.~Kopley, R.~Jewett,
  J.~Pernillo, C.~Tan, and A.~Montijo, \enquote{A 20 GS/s 8 b ADC with a 1 MB
  memory in 0.18 /spl mu/m CMOS,} in \emph{IEEE International Solid-State
  Circuits Conference Digest}, pp. 318--496 vol.1 (2003).

\bibitem{GuptaISSCC2006}
S.~Gupta, M.~Choi, M.~Inerfield, and J.~Wang, \enquote{A 1GS/s 11b
  Time-Interleaved ADC in 0.13 um CMOS,} Proc. IEEE ISSCC Dig. Tech. Papers pp.
  576--577 (2006).

\bibitem{TEKOSC}
\enquote{Tektronix product information,}
  \urlprefix\url{http://www.tek.com/products/oscilloscopes/selection\_chart.ht%
ml.}

\bibitem{LECROYOSC}
\enquote{LeCroy Product Information,}
  \urlprefix\url{http://www.lecroy.com/tm/products/default.asp.}

\bibitem{BartolomeTI2005}
E.~Bartolome, V.~Mishra, G.~Dutta, and D.~Smith, \enquote{Clocking high-speed
  data converters,} TI Analog Applications Journal  (2005).

\bibitem{Chen2007}
J.~Chen, J.~Sickler, E.~Ippen, and F.~K{\"a}rtner, \enquote{High repetition
  rate, low jitter, low intensity noise, fundamentally mode-locked 167 fs
  soliton Er-fiber laser,} Optics Letters \textbf{32}(11), 1566--1568 (2007).

\bibitem{Zanchi2005}
A.~Zanchi and F.~Tsay, \enquote{A 16-bit 65-MS/s 3.3-V pipeline ADC core in
  SiGe BiCMOS with 78-dB SNR and 180-fs jitter,} IEEE Journal of Solid-State
  Circuits \textbf{40}(6), 1225--1237 (2005).

\bibitem{GuptaOL2008}
S.~Gupta and B.~Jalali, \enquote{Time-warp correction and calibration in
  photonic time-stretch analog-to-digital converter,} Opt. Lett.
  \textbf{33}(22), 2674--2676 (2008).
  \urlprefix\url{http://ol.osa.org/abstract.cfm?URI=ol-33-22-2674}.

\bibitem{OFSDCF}
\enquote{OFS dispersion compensation fiber product datasheet,}
  \urlprefix\url{http://www.specialityphotonics.com/pdf/products
  /speciality/dispersion/HFDK-C.pdf.}

\bibitem{GuptaJLT2007}
S.~Gupta, G.~C. Valley, and B.~Jalali, \enquote{Distortion Cancellation in
  Time-Stretch Analog-to-Digital Converter,} J. Lightwave Technol.
  \textbf{25}(12), 3716--3721 (2007).
  \urlprefix\url{http://jlt.osa.org/abstract.cfm?URI=JLT-25-12-3716}.

\bibitem{JuodawlkisMTT2001}
P.~W. Juodawlkis, J.~C. Twichell, G.~E. Betts, J.~J. Hargreaves, R.~D. Younger,
  J.~L. Wasserman, F.~J. O'Donnell, K.~G. Ray, and R.~C. Williamson,
  \enquote{Optically sampled analog-to-digital converters,} IEEE Transactions
  on Microwave Theory and Techniques \textbf{49}(10), 1840--1853 (2001).

\bibitem{Ouellette1994}
F.~Ouellette, J.-F. Cliche, and S.~Gagnon, \enquote{All-fiber devices for
  chromatic dispersion compensation based on chirped distributed resonant
  coupling,} Journal ofLightwave Technology \textbf{12}(10), 1728--1738 (1994).

\bibitem{Conway2008}
J.~A. Conway, G.~A. Sefler, J.~T. Chou, and G.~C. Valley, \enquote{{Phase
  ripple correction: theory and application},} Optics Letters \textbf{33}(10),
  1108--1110 (2008).

\bibitem{Stigwall2007}
J.~Stigwall and S.~Galt, \enquote{Signal Reconstruction by Phase Retrieval and
  Optical Backpropagation in Phase-Diverse Photonic Time-Stretch Systems,}
  Journal of Lightwave Technology \textbf{25}(10), 3017--3027 (2007).

\bibitem{ChoJLT2004}
H.~Cho, P.~Kapur, and K.~Saraswat, \enquote{{Power Comparison Between
  High-Speed Electrical and Optical Interconnects for Interchip
  Communication},} Journal of Lightwave Technology \textbf{22}(9), 2021--2033
  (2004).

\bibitem{McFadden2006}
M.~J. McFadden, M.~Iqbal, T.~Dillon, R.~Nair, T.~Gu, D.~W. Prather, and M.~W.
  Haney, \enquote{Multiscale free-space optical interconnects for intrachip
  global communication: motivation, analysis, and experimental validation,}
  Applied Optics \textbf{45}(25), 6358--6366 (2006).

\end{thebibliography}
\end{document}